\documentclass[traditabstract]{aa}
\usepackage{amssymb,amsmath,amsfonts}
\usepackage{graphicx,graphics}

\newcommand{\beq}{\begin{equation}}
\newcommand{\eeq}{\end{equation}}
\newcommand{\bea}{\begin{eqnarray}}
\newcommand{\eea}{\end{eqnarray}}
\newcommand{\ks}{\kappa_{\rm S} }
\newcommand{\kt}{\kappa_{\rm T} }
\newcommand{\ras}{\mathrm{Ra_*} }

\newcommand{\rast}{Ra$_*$ }

\newcommand{\nut}{{\mathrm{Nu_T}} }
\newcommand{\nutt}{Nu$_{\rm T}$ }
\newcommand{\nus}{{\mathrm{Nu_S}} }
\newcommand{\nust}{Nu$_{\rm S}$ }
\begin{document}
\title{Semiconvection}
%\titlerunning{to be set}

\author{F.\ Zaussinger\inst{\ref{inst1},\ref{inst2}}, H.C.\ Spruit\inst{\ref{inst1}}}
\authorrunning{F.\ Zaussinger \& H.C.\ Spruit}

\offprints{\\ H.\ Spruit, \email{henk@mpa-garching.mpg.de}}

\institute{
  Max-Planck-Institut f\"{u}r Astrophysik,
  Karl-Schwarzschild-Str.\ 1,
  D-85748 Garching, Germany \label{inst1}
\and 
Faculty of Mathematics, 
University of Vienna, 
Nordbergstra\ss e 15, 
A-1090 Vienna, Austria \label{inst2}
}
\date{\today}

\abstract{
A grid of numerical simulations of double-diffusive convection is presented for astrophysical conditions. As in laboratory and geophysical cases convection takes place in a layered form. A translation between the astrophysical fluid mechanics and incompressible (Boussinesq) approximation is given, valid for thin layers.  Its validity is checked by comparison of the results of fully compressible and Boussinesq simulations of semiconvection. A fitting formula is given for the superadiabatic gradient as a function of this parameter. The superadiabaticity depends on the thickness $d$ of the double diffusive layers, for which no good theory is available, but the effective He-diffusion coefficient is nearly independent of $d$.  For a fiducial main sequence model (15 $M_\odot$) the inferred mixing time scale is of the order $10^{10}$ yr. 
}

\maketitle

\section{Introduction}
In models of stellar structure, situations are found where the heavier products of nuclear burning provide stability to a zone which otherwise would be unstable to convective overturning. Such a zone, or part of it, would become convective if something managed to mix its composition (R.J.\ Tayler, 1953). The question whether such a zone should be treated as if it were mixed or not has become known as the {\em semiconvection} problem. Answers to this question differ substantially. In practice, recipes are used containing a free parameter that allows the degree of mixing to be varied. Calculations in which such a parameter is adjusted to match observations are then called `with semiconvection'.  Commonly used prescriptions are those of Langer (1985) and Maeder (1997).

The presence of a semiconvective zone has only a minor effect on the thermal structure of the star. The assumed amount of mixing of composition is critical, however, because the evolution of the star is sensitive to the precise distribution of products of nuclear burning with depth in the star. The main goal of a theory for semiconvection is thus a good determination of the rate of mixing. From the perspective of the stellar evolutionist, the theory would ideally provide a formula for the rates of mixing and energy transport (the effective diffusivities), as functions of local thermodynamic state and composition, and their gradients.

In Spruit (1992, hereafter S92) such formulas were derived, adapting the known physics of \textit{double-diffusive convection} to the case of a stellar interior. In the following, numerical simulations are used to measure mixing rates and their dependence on astrophysical conditions. They are compared with S92, and used as a basis for updated fitting formulas for the mixing rates in semiconvective zones of stars.

\section{Semiconvection and double-diffusive convection}
Situations where a fluid is stabilized by the density gradient due to a dissolved heavy constituent occur in nature. An example is convection under the arctic ice sheet (cooled from above, stabilized by the salts dissolved in the sea water). Particularly intensively studied are East-African rift lakes (lakes Kivu, Nyos and Mounon, cf.\ Schmid et al. 2010). These are heated from below by volcanic activity, which also is a source of dissolved gases (carbon dioxide and methane, hereafter the `solute'). Their density stratification is stabilized against convection by the stable gradient resulting from the weight of the carbon dioxide. Efforts to prevent catastrophic release of carbon dioxide (Lake Nyos, e.g.\ Sigvaldason 1989) or commercial exploitation of methane (Lake Kivu, Nayar 2009) have led to extensive study of the fluid flows, heat flux and mixing rates in these natural double-diffusive systems.

The gradients in temperature and solute in these lakes are observed to be `stepped': consisting of a stack of thin layers (decimeters to decameters). Inside a layer, overturning convection keeps the composition nearly uniform, with stable jumps in temperature and composition separating the layers. The physics involved is easily reproduced under controlled laboratory conditions (Turner 1985).\footnote{Also on a coffee table. A latte macchiato in a tall glass often shows the effect nicely. After the coffee is added to the milk, a stably stratified gradient of milk/coffee mix develops (showing internal gravity waves in the form of a sloshing motion with a period of a few seconds). After about a minute, the initially smooth gradient starts dividing into thin (a few mm) layers, visible at low contrast. In the course of several minutes these merge into a smaller number of more clearly defined layers.} The layers are very long-lived:  of the order of months or more in the geophysical examples mentioned,   orders of magnitude  longer than the convective turnover times inside the layers.

In the stable steps between the layers the transport of the stabilizing solute takes place by diffusion instead of convection. This strongly limits the effective transport of solute through the double-diffusive stack of layers. Residence times on the of order 1000 yrs are inferred for the solutes in lake Kivu, for example (Schmid et al. 2010). This is 8 orders of magnitude longer than the convective turnover times in these layers. The transport of heat is also strongly reduced, this is exploited for heat storage in solar ponds (cf. Lu and Swift, 2001). 

Theoretically, the observed layered nature of double-diffusive systems is well understood. Central to this understanding is the fact that linear stability analysis does not provide relevant clues to their behavior, because the double-diffusive case of thermal convection stabilized by a slowly-diffusing solute is {\em subcritical}. That is, stable overturning flows occur at a temperature gradient {\em below} the onset of linear instability. 

Linear instability predicts internal gravity modes to set in above some critical value of the temperature gradient, growing in amplitude by the effect of thermal diffusion, the so-called Kato oscillations (Kato 1966). Such oscillations (cf. movie at Fig.\ \ref{kato}) transport a negligible amount of heat or solute, compared with overturning motions of the same amplitude. For this reason alone, linear stability arguments cannot be used for useful estimates of the mixing rate in semiconvective zones. More important is the subcritical nature of double-diffusive convection. Proctor (1981) shows that, in the limit of vanishing diffusivity of the solute, the layered form of convection can occur whenever the temperature gradient is unstable to convection (i.e. `according to Schwarzschild'), irrespective of the strength of the stabilizing component.

The reason for this subcritical behavior can be understood with an energy consideration. The amount of energy it takes to overturn a layer of thickness $d$ against a stable gradient scales as $d^2$. Per unit mass the expense in initial energy  needed to put the system into its finite-amplitude, layered state thus vanishes as $d$, down to the value where energy loss by viscous damping stabilizes the system. This minimum is set by the critical Rayleigh number for ordinary convection in a layer of thickness $d$. A small initial perturbation, or perhaps an initial Kato oscillation, is sufficient to provide the energy for overturning into thin layers. Once established, this layered state is a stable form of convection. This agrees with the observation that in laboratory experiments and geophysical systems like lake Kivu mentioned above the layering first sets at a small thickness (cf. the latte macchiato experiment in the footnote above).

\subsection{Semiconvection}
It is sometimes argued that the geophysical and laboratory examples of double diffusive convection cannot be applied to a stellar interior because of different physics. At the level of physics  in evidence in  current recipes for semiconvection, however, there is no distinction between these cases. The equations of fluid dynamics used are the same. In contrast with astrophysics, an incompressible approximation is often used in geophysics, but this is not mandatory. The compressibility of water can be, and has been included widely in geophysical fluid mechanics.

The fluid is described by the thermodynamic variables defining its local state (e.g. pressure and density), the gradients of these with depth, the transport properties that are determined by the thermodynamic state (viscosity $\nu$, thermal diffusivity $\kt$ and solute diffusivity $\ks$), the acceleration of gravity, and the equation of state. Taken together, these quantities form a large parameter space, and it might be concluded that realistic numerical simulations of semiconvection would have to be done for individual zones in individual stellar models. 

The equations of fluid dynamics have symmetries, however, so that the independent degrees of freedom are far fewer. They can be represented by 5 dimensionless parameters: a Rayleigh number Ra, the layering thickness $\epsilon=d/H$ in units of the pressure scale height $H_P$, the Prandtl number Pr$=\nu/\kt$, the Lewis number $\mathrm{Le}=\ks/\kt$, and a density ratio $R_{\rho}$ which measures the ratio of stabilizing (solute) to destabilizing thermal gradient. The behavior of semiconvection at any point in a star can be defined in terms of these parameters.

This is discussed in more detail in section \ref{Bouss}, where it is also shown how in the limit $\epsilon \ll 1$ the astrophysical problem can be translated into an equivalent incompressible (Boussinesq) problem.  In section \ref{compar} we verify this with a direct comparison between results from fully compressible and Boussinesq simulations. In this limit, the $\epsilon$ disappears from the problem, reducing the number of parameters from 5 to 4. 

By a fortunate coincidence, it turns out  that the results are effectively independent of $\mathrm{Pr}$, as long as it is small. Which is in fact the case in a stellar interior. This further reduces the number of independent parameters to only 3. Since measurement of the mixing rate in each individual case does not require a very expensive simulation, a significant volume of astrophysically relevant parameter space can be covered.

The semiconvective zone in a stellar model is sufficiently limited in extent and consequences that the overall structure of the star does not depend much on the way semiconvection is calculated. The additional diffusion of Helium by semiconvective mixing can be important for later evolutionary stages, but during the semiconvective phase itself the thermal structure of the star is not affected much. The consequence of this is that in contrast with to laboratory and geophysical situations, in a stellar model the {\em heat flux} $F$ can be considered as known, rather than the temperature gradient. Since the radiative contribution $F_{\rm r}$ to the heat flux is known to good approximation from the thermal structure of the star, the convective heat flux $F_{\rm c}=F-F_{\rm r}$ transported by semiconvection is then also known. What is not yet known is the efficiency of convection: i.e. how close to the adiabatic gradient the mean thermal gradient will turn out to be as a result of semiconvective transport.
\subsection{Layered convection}
Asymptotically at high Rayleigh number, the convection inside each layer of the semiconvective stack is driven entirely by boundary layers at the top and bottom steps of the layer, much like in laboratory convection in a box. Except for these thin boundary layers, the (horizontal averages of) entropy and composition are almost uniform inside the layers. 

Under the conditions in a stellar interior, the thermal diffusivity is much larger than the diffusivity of the solute (Helium in Hydrogen, say). The thickness of the solute boundary layers is then much smaller than the thermal boundary layers, and the convective flow almost the same as in the absence of a stabilizing solute. Under these asymptotic conditions, the dependence of heat flux on Rayleigh number can be taken from a simple estimate, as was done in S92, or a laboratory result can be used (e.g.\  Niemela et al. 2000). With the convective flow thus known, the flux of solute can be calculated as well. 

One of the consequences of such an asymptotic model is the prediction that the effective diffusivity of the solute scales with the square root of its microscopic value. This is the well-known relation derived in various ways for double-diffusive convection (cf. Turner 1985), and verified in laboratory experiments. It will show up again in the numerical experiments reported below.

\subsubsection{Layer thickness}
The main uncertainty in applying results on double-diffusive convection to stars (or any other double-diffusive situation) is the poorly known physics which determines the thickness of the individual layers in the stack. The layer thickness thus remains a free parameter of the problem. 

\subsection{Boundary layers}

\label{bound}
Each of the layers in the stack is separated from the next by a step in composition and temperature. The step consists of three nested boundary layers, for temperature, composition, and flow speed, called the {\em thermal}, {\em solute} and {\em viscous} boundary layers respectively. Their thicknesses are governed by the thermal diffusivity $\kt$, solute diffusivity $\ks$, and (kinematic) viscosity $\nu$. The highest diffusivity (thermal) has the widest boundary layer. For astrophysical conditions, $\kt\gg\ks,\nu$, the solute and viscous sublayers are so thin compared with the thermal boundary layer that their influence on the convective flow pattern is negligible.
The viscous sublayer is different from the other two. Whereas temperature and solute change rapidly across the step, the parallel flow component varies smoothly across the step. The stable step in composition confines the flow to the layer. Apart from of deformations of the step in the form of stable interface waves,  the vertical velocity component vanishes at the step. 

Summarizing the above, in the astrophysical limit $\kt\gg\ks,\nu$, the flow in the layer behaves to a good approximation like convection between horizontal plates, with free slip conditions at these boundaries. This allows us to make estimates of the thickness of the boundary layers. The boundary layers determine the fluxes of heat and solute through the layer, i.e. the numbers we are interested in. The need to properly resolve them numerically determines the required number of grid points in the vertical direction.

Let $\tau_{\rm c}$ be the time the overturning flow is in contact with the boundary, before descending/ascending into the interior of the layer. Diffusion of heat through the step over this time sets the thickness $\delta_{\rm T}$ of the thermal boundary layer:

%In the context of the fluid column height d, the thermal boundary layer thickness $\delta_{\rm T}$ is an 
%unknown parameter. With the help of the mixing length theory (MLT) a correlation between the 
%Rayleigh number and the thermal diffusion coefficient can be obtained.
%
%We define the thickness of the thermal boundary layer in a semi-convection zone as, 
%
\beq
\delta_{\rm T} \approx (\kt\tau_{\rm c})^{1/2}.
\label{deltaT-deriv}
\eeq
%To calculate $\delta_{\rm T}$ the thermal diffusion coefficient $\kappa_{\rm T}$ and the convective time scale $\tau_{\rm c}$ has to be determined. The convective cell time scale $\tau_{\rm c}$ is obtained by
Assuming that the convective cells driven by the boundary layers have a horizontal width of the order of the thickness $d$ of the layer, $\tau_{\rm c}$ is of the order of the convective overturning time, $\tau_c \approx d/v_{\rm c}$, 
where the convective velocity $v_{\rm c}$ can be estimated by the mixing-length expression, 
\beq
v^2_c = \frac{\eta}{8}g\, (\nabla-\nabla_{\rm a})\frac{d^2}{H}.
\label{velconv}
\eeq
[Standard notation: pressure scale height $H_P=1/({\rm d}\ln P/{\rm d}z)$, $\nabla={\rm d}\ln T/{\rm d}\ln P$, $\nabla_{\rm a}$ the corresponding  adiabatic gradient, and $\eta=-\partial \ln \rho/\partial \ln T\vert_P$ is a number of order unity (called $\delta$ in Kippenhahn \& Weigert 1990, p39), $\eta=1$ in a fully ionized ideal gas]. 
Define a {\em modified Rayleigh number} $\ras$ by
\beq \ras=\mathrm{Pr\, Ra}. \eeq
In astrophysical notation, it is given by
\beq
\ras = \frac{g}{H}(\nabla-\nabla_{\rm a})f(\beta)\frac{d^4}{\kappa^2_{\rm T}}\sim {v_{\rm c}^2d^2\over \kt^2},\label{ras}
\eeq
where $g$ is the acceleration of gravity, and $f(\beta)$ a factor depending on the ratio, $\beta = P_{\rm g}/P$, of gas pressure $P_{\rm g}$ to total (gas plus radiation) pressure,  $P=P_{\rm g}+P_{\rm rad}$. $\ras$ thus contains only the thermal diffusivity, not the viscosity. Except for Rayleigh numbers close to the critical value Ra$_{\rm c}$ for instability to convection, it is the quantity that determines the heat flux (Nusselt number), in the limit of low Prandtl number. The simulations below will also address conditions close to marginal, but $\ras$ is the meaningful number for translating them to astrophysical application, where Ra $\gg {\rm Ra}_{\rm c}$. Using equations \eqref{velconv} and \eqref{deltaT-deriv},
\beq
\delta_{\rm T}^4 \approx \kappa^2_{\rm T} \tau_{\rm c}^2\approx \frac{8}{\eta}f\frac{d^4}{\ras}.
\eeq
Since $8 f/\eta$ is a numerical factor of order unity, our estimate of the thermal boundary layer thickness is
\beq \delta_\mathrm{T}/d\approx\mathrm{Ra}_*^{-1/4}.\label{delt}\eeq

The {\em Nusselt number}, the ratio of flux in the presence of the flow to the flux in the absence of a flow is then just the ratio of the gradient in the boundary layer to the average gradient between top and bottom. The expected dependence on \rast is thus, with (\ref{delt}):
\beq \mathrm{Nu_T}\approx d/\delta_{\rm T}=\ras^{1/4}.\eeq

The thickness of the solute (Helium-) boundary layer can be estimated by the same reasoning,
\beq
\delta_{\rm S} \approx  (\ks \tau_{\rm c})^{1/2} = \mathrm{Le}^{1/2}(\kt\tau_{\rm c})^{1/2} =\mathrm {Le}^{1/2}\delta_{\rm T},\label{dels}
\eeq
where $\mathrm{Le}$ is the Lewis number  $\mathrm{Le} = \ks/\kt$ as before.
Being the thinnest boundary layer in the problem, $\delta_{\rm S}$ determines the numerical resolution needed. Since the fine structure in the interior of the layer largely consist of plumes `peeled off' from the boundaries, the same resolution is needed in the interior of the layer as well, and there is no need or justification for using non-uniform grids.

\subsection{Density ratio, mixing rate}

The problem also depends on the relative strength of the stabilizing solute gradient relative to the destabilizing thermal gradient. This can be measured in terms of the thermal and solute buoyancy frequencies $N_{\rm T}$, $N_{\rm S}$:
\beq N_{\rm T}^2={g\over H}(\nabla_{\rm a}-\nabla),\eeq
\beq N_{\rm S}^2=g\,{\rm d}\ln\mu/{\rm d} z,\eeq
where $\mu$ is the `mean molecular weight'. The {\em density ratio} $R_\rho$:
\beq R_\rho\equiv-N_{\rm S}^2/N_{\rm T}^2\label{rrho}\eeq
is then the dimensionless measure we will use for the relative strength of the stabilizing solute gradient.  The minus sign is added for convenience to the definition, since $N_{\rm T}^2<0$ for a convectively unstable stratification. Semiconvection,  i.e. a stratification `between Schwarzschild and Ledoux' then corresponds to $R_\rho>1$.

The amount of solute transported across the layer is limited by the fact that convective plumes develop only from material with a net buoyancy of the unstable sign. This limits the density contrast of the stable solute carried to a fraction $1/R_\rho$ of the density contrast across the layer. With the same reasoning used above for the heat flux, this gives a relation between the solute Nusselt number \nust and \nutt:
\beq \nus={\rm Le}^{-1/2}\nut/R_\rho\label{nusnut0}.\eeq
(cf. S92). Solute with a higher density contrast is not mixed into the convective flow: it forms a {\em stagnant} region around the interface. This region has a width 

\beq d_{\rm s}\approx\delta_{\rm S}R_\rho, \eeq 
which decreases the solute gradient and thereby the solute diffusion rate at the interface. The effect is equivalent to the reduction of \nust by the factor $R_\rho$ in (\ref{nusnut0}).

In the above (as in S92), we have implicitly assumed that $R_\rho$, while $>1$, is still low enough that the stagnant region is thinner than the thermal boundary layer thickness:
 \beq d_{\rm S} < \delta_{\rm T}, \eeq
 so that it does not interfere with the convective flow. An improvement for the more general case  yields
 \beq  \nus=(1+{\rm Le}^{-1/2}/R_\rho)\nut\label{nusnut1}.\eeq
As discussed in section \ref{stars}, the Lewis number is so small in a stellar interior that this correction is not likely to become important, and estimate (\ref{nusnut0}) is still good enough.
The heat flux can be considered as fixed by the stellar structure, so \nutt is fixed and Eq. (\ref{nusnut0}) also fixes the mixing rate, largely independent of the unknown layer thickness $d$. 
\section{Boussinesq limit: thin layers} 
\label{Bouss}

In a stellar interior, radiation carries a heat flux even in a convectively neutral stratification. This requires some care when making the connection with the Boussinesq case.

\subsection{heat flux}
Let $F^{\rm s}_{\rm r}$, $F^{\rm s}_{\rm c}$ be the radiative and hydrodynamic (semiconvective) contributions to the heat flux in the star, where the superscript $^{\rm s}$ designates quantities under the compressible-fluid conditions in a star, while  $^{\rm B}$ will be used for the corresponding quantities in the Boussinesq (incompressible) model.
The radiative heat flux is proportional to the temperature gradient $\nabla$:
\beq
F^{\rm s}_{\rm r}= q\nabla= q\nabla_{\rm a} + q(\nabla-\nabla_{\rm a})\equiv F^{\rm s}_{\rm ra}+F^{\rm s}_{\rm rs},
\eeq
where $q$ is a constant depending on the local thermodynamic state, and $F^{\rm s}_{\rm ra}$, $F^{\rm s}_{\rm rs}$ are the contributions to the \underline{r}adiative heat flux of the \underline{a}diabatic and the \underline{s}uperadiabatic parts of the mean temperature gradient. In the Boussinesq model, the contribution $F_{\rm ra}$ is absent: the convective and radiative heat fluxes $F^{\rm B}_{\rm c}$, $F^{\rm B}_{\rm r}$ are governed by the same temperature gradient. Related to this, the Boussinesq model has one fewer parameter: the pressure scale height $H$. Because of this difference, the  heat flux in the stellar (compressible) model cannot be compared directly with  the heat flux in an incompressible model. Instead, the ratio $f^{\rm B}$ of convective to radiative heat flux in the Boussinesq model  is to be identified with the ratio $ f^{\rm s}=F^{\rm s}_{\rm c}/F^{\rm s}_{\rm rs}$ of the convective flux to the {\em superadiabatic component} of the radiative flux  in the star, in the limit $H\rightarrow\infty$. If $\epsilon=d/H$, where $d$ is the thickness of the convecting layer, then for a given value of the modified Rayleigh number Ra$_*$ (cf.\ Massaguer \& Zahn 1980):

\beq f^{\rm B} ({\rm Ra_*};\{p\})=f^{\rm s}({\rm Ra_*},\epsilon\downarrow 0;\{p\}),\label{ident}\eeq
where $\{p\}$ stands for the other dimensionless parameters of the problem: $\rm Pr$, $\rm Le$ and $R_\rho$. The Boussinesq model thus is the limit  $\epsilon \downarrow 0$ taken {\em at fixed} $\ras$. If the semiconvective layer thickness $\epsilon$ is `sufficiently small' (in some sense be verified by numerical tests), the compressible case can be compared directly with the Boussinesq model, in the sense of eq.\ (\ref{ident}). 

As mentioned in the introduction, this makes the semiconvection problem particularly amenable to numerical simulation. In contrast with a convective stellar envelope, for example, with its many scale heights to be covered, the layered nature of double diffusive convection puts it in a parameter range that is much more accessible with realistic ab initio calculations.

To complete the Boussinesq case, we still need a translation of the density ratio $R_\rho$. With eq. (\ref{rrho}) this can be done uniquely in terms of the buoyancy frequencies, see eqs.\ (\ref{NTB},\ref{NSB}) below.

\section{Numerical simulations}
All equations were calculated on a 2D rectilinear Cartesian grid in terms of finite differences. The set of equations are implemented into the ANTARES software framework (Muthsam et al.\ 2010). Advective currents are solved by a weighted essentially non-oscillatory finite volume scheme in fifth order (Shu \& Osher 1988), the physical diffusion is handled by a fourth-order finite difference discretization. A second order total variation diminishing scheme as time integrator is chosen. To avoid odd-even decoupling, a MAC grid, which locates vector variables at cell faces and scalar variables at cell centres, is used.

For a detailed description of the numerical solution of the binary mixture equations presented here see Zaussinger (2011).

\subsection{Boussinesq approximation for a binary mixture}
In the Boussinesq approximation the continuity equation is reduced to that of an incompressible fluid,

\begin{equation}
\frac{{\mathrm d} \rho}{{\mathrm d} t} =  0, \label{BA1}
\end{equation}
such that in the equation of motion the density  is taken to be a constant $\rho_0$,

\begin{equation}
\frac{{\mathrm d} \vec u}{{\mathrm d}t} = -{1\over\rho_0}\nabla P + \left ( - \alpha_{\rm T}\Theta + \alpha_{\rm S}S \right ) {\vec g} + \nabla \cdot (\nu \nabla \vec u),  \label{BA2}
\end{equation}
where the `expansion coefficients' $\alpha_{\rm T}, \alpha_{\rm S}$ describe the density effects   of variations $\Theta$, $S$ in temperature and solute, assumed small compared with the mean temperature $\bar\Theta$ and solute $\bar S$. They are given by advection-diffusion equations: 

\begin{equation}
\frac{{\mathrm d} \Theta}{{\mathrm d}t }= \vec u \cdot \nabla \bar{\Theta} +  \nabla \cdot  (\nabla \kappa_{\rm T} \Theta),  \label{BA3}
\end{equation}

\begin{equation}
\frac{{\mathrm d} S}{{\mathrm d}t}= \vec u \cdot \nabla \bar S + \nabla \cdot  (\kappa_{\rm S} \nabla S),  \label{BA4}
\end{equation}
where $\nu$ is the kinematic viscosity. The mean gradients $\nabla\bar\Theta$, $\nabla\bar S$ can be expressed more usefully in terms of the buoyancy frequencies:
\beq N_{\rm T}^2=\alpha_{\rm T}\,{\bf g}\cdot\nabla\bar\Theta,\label{NTB}\eeq
\beq N_{\rm S}^2=\alpha_{\rm S}\,{\bf g}\cdot\nabla\bar S,\label{NSB}\eeq
The density ratio $R_\rho$ is then defined as in the compressible case, eq.\ (\ref{rrho}).

%The pressure $P$, the potential temperature $\Theta$ and the solute $S$ are expanded in terms of the Boussinesq approximation as a sum of the basic profile $\bar P$,$\bar{\Theta}$, $\bar S$ and the fluctuation $P^\prime$, $\Theta^\prime$, $S^\prime$. 

The numerical algorithm solving this set of equations is based on a semi-implicit scheme. Intermediate values for the velocity field $\vec u^*$ are calculated explicitly from the equations of motion. By the nature of the incompressible equations, the pressure update is done implicitly, by solving a Poisson equation:

\begin{equation}
\Delta P = \frac{\rho_0}{\Delta t} (\nabla \cdot \vec u^*)  \label{Poisson}
\end{equation}

The resulting pressure $\rm P$ leads to the required divergence free velocity field at the new time step $n+1$.

\begin{equation}
\vec u^{n+1} = \vec u^* - \frac{\Delta t}{\rho_0} \nabla P.  \label{update}
\end{equation}

\subsection{Compressible fluid equations for a binary mixture}
Verification of the Boussinesq results  has been done with simulations of the fully explicit compressible fluid equations. The fluid is assumed to be an ideal gas, which is a good approximation to a binary gas mixture of our interest. These are the continuity equation

\begin{equation}
\frac{\partial \rho}{\partial t} + \nabla \cdot (\rho \vec u )=0
\end{equation}

the partial density equation
\begin{equation}
\frac{\partial (\rho c)}{\partial  t} + \nabla \cdot (\rho  c  \vec u )= \nabla \cdot (\rho \kappa_c \nabla c)
\end{equation}
 
the momentum equation
\begin{equation}
\frac{\partial (\rho \vec u)}{\partial t} + \nabla \cdot (\rho \vec u\vec u + P \underline{\rm I} )= \rho  g_z + \nabla \cdot \tau
\end{equation}
 
the total energy equation
\begin{equation}
\frac{\partial e}{\partial t} + \nabla \cdot [ \vec u( {e + P}) ]=  \rho (\vec  g \vec u) + \nabla \cdot (\vec u \tau)
\end{equation}

and the equation of state

\begin{equation}
P = \frac{\mathcal{R} \rho T}{\mu (c)}
\end{equation}

where $c$ is the solute mass fraction, $\kappa_c$ is the solute diffusion coefficient, $\tau$ is the viscous stress tensor and $\mu = \mu(c)$ the mean molecular weight. 

\subsection{Units, boundary and initial conditions}
\label{bc}

As unit of length we use, for the Boussinesq cases the layer thickness $d$, for the compressible calculations the pressure scale height $H$.  The nominal convective turnover time $d/v_{\rm c}$ from (\ref{velconv}) is used as unit of time. As units of temperature (potential temperature $\Theta$ in the compressible case) we use $1/\alpha_{\rm T}$, for solute concentration $1/\alpha_{\rm S}$. The density ratio then becomes $R_\rho=\nabla \bar S/\nabla \bar T$. The Boussinesq equations are invariant to arbitrary additive constants in temperature and solute. We set these such that $T$ and $S$ are zero at the top boundary.

Because of the symmetries of the problem, there are fewer independent parameters than physical variables describing it. Hence some of the physical quantities appearing in the problem can be set to unity. We choose for these: the temperature difference between top and bottom of the layer, the density at the bottom of the layer and the acceleration of gravity. The Rayleigh number Ra$_*$ is then controlled through the thermal diffusivity $\kappa_{\rm T}$, the solute difference between top and bottom though the density ratio $R_\rho$, and the solute diffusivity through the Lewis number Le. 

Most of the calculations were done in a box simulating a single layer from the double-diffusive stack, so the top and bottom boundaries coincide with the steps between layers. This ignores the distortions of the interfaces by surface waves, but since the essence of the double layering phenomenon is that the transport across the interface is by diffusion, this is not expected to make a big difference. To check that this is indeed the case, a smaller set of simulations was done in which a step is present inside the volume (section \ref{double}). The vertical boundary conditions are thus taken to be impermeable and stress-free. In the horizontal direction periodic conditions are used.   
\begin{eqnarray}
u_z &= 0 \qquad z =& 0, 1 \\
\frac{\partial u_x}{\partial z}&=0 \qquad z =& 0, 1 \\
S &=R_\rho \qquad z =& 0 \\
T &=1 \qquad z =& 0 \\
S& = 0 \qquad z =& 1 \\
 T&= 0 \qquad z =& 1
\end{eqnarray}      
As initial condition the stratification of temperature and solute was taken to be horizontally uniform with either a linear gradient between the values at top and bottom (the `linear' case below), or something approximating the boundary layer structure expected of the final state (the `step' case). Small initial random perturbations are applied on the solute field. 

The numerical algorithm for setting up the initial conditions for the compressible fluid equations is an extension of the procedure presented in (Muthsam et al.\ 1995,\ 1999).

\subsection{Numerical setup}

Being the thinnest boundary layer in the problem, $\delta_{\rm S}$ determines the numerical resolution needed near the boundaries. Since the fine structure in the interior of the layer largely consist of boundary layers `peeled off' from the boundaries, the same resolution is needed in the interior of the layer as well, and there is no need or justification for using non-uniform grids.

Using  Ra$_*=1.6 \times 10^5$ and a Lewis number of $\rm Le = 0.1$ would lead to solute boundary thickness of $1.6\%$ of the layer thickness $d$. With the high-order spatial discretization used, a minimum resolution of $3$ points is then needed in the boundary layers, which translates to  $200$ points in vertical direction for this case. Convergence tests showed that a numerical resolution of $300$ points was sufficient for  the range in Lewis number considered. Most of the calculations were done with a horizontal-to-vertical aspect ratio of 2:1. A few tests with different ratios showed that this choice does not affect the measured Nusselt numbers significantly (section \ref{aspect})

\section{Results of numerical simulations}

The numerical results presented here are based on about $100$ numerical simulations done in the Boussinesq approximation and about $20$ simulations performed with the fully compressible code. 

\begin{figure}
\begin{center}
\includegraphics[width=0.9\hsize]{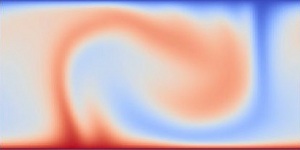}\\
\vspace{0.4\baselineskip}
\includegraphics[width=0.9\hsize]{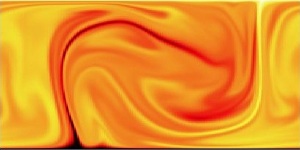}
\caption{Flow structure in a double diffusive layer. The temperature field (top) is more diffuse than the solute (`Helium') concentration, as a result of the high thermal diffusivity . }
\label{sample}
\end{center}
\end{figure}

An example is shown in Fig.\ \ref{sample}, with $\rm Pr=0.1$, Ra$_*=5\times10^5$, $\rm Le=0.01$, R$_\rho$=1.15. It shows the key characteristic of double diffusive convection:  the  boundary layers and the plumes of the solute are narrower than those of temperature, on account of the low solute diffusivity. The flux of solute carried by the velocity field is correspondingly lower, and vanishes in the limit $\mathrm{Le}\rightarrow 0$ (cf. eq. \ref{nusnut}).

%The heat flux is conventionally measured by the thermal Nusselt number
%\beq \nut=F/F_0,\eeq
%where $F$ is the (time averaged) heat flux and $F_0$ is the heat flux in the absence of a flow (for the Boussinesq case. In the compressible case, $F_0$ is the radiative heat flux carried by the superadiabatic part of the temperature gradient, see section \ref{Bouss}). The solute Nusselt number \nust is defined analogously. 
For comparison with the numerical results, we use the theoretical estimates in \ref{bound} .
For large Nu$_{\rm T}$, Nu$_{\rm S}$, the analysis in \ref{bound} predicts a classical double-diffusive square root dependence on the ratio of diffusivities:
\beq \nus\approx (1+\mathrm{Le}^{-1/2}/R_\rho)\,\nut\label{nut} \eeq
which is observed in laboratory experiments. This estimate assumes the boundary layers to be thin compared with the layer thickness. An estimate which is slightly better at low Nusselt numbers is obtained by noting that the convective flux $F-F_0$ is a more relevant measure of the double diffusive transport efficiency than the total flux $F$; this yields:
\beq (\nus-1)\approx(1+ \mathrm{Le}^{-1/2}/R_\rho)\,(\nut-1). \label{nusnut}\eeq
We use these estimates for comparison with the numerical results.

\subsection{Dependence on \rast and $R_\rho$}

\label{lerho}
\begin{figure}
\begin{center}
\includegraphics[width=\hsize]{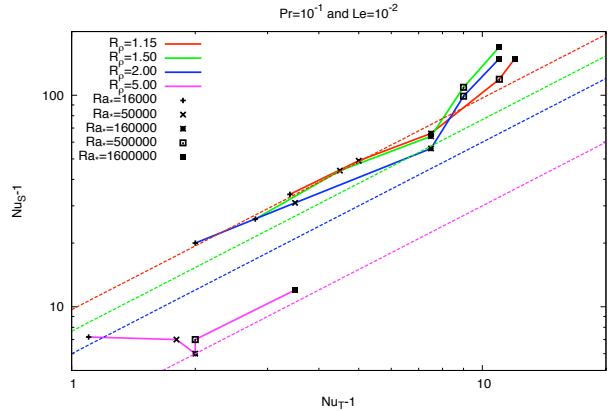}
\caption{The relation between Nu$_T$ and Nu$_S$ for $\rm Pr=10^{-1}$ and $\rm Le=10^{-2}$. Parameter along the solid curves is the Rayleigh number Ra$_*$, colors indicate density ratio $R_\rho$. Dotted lines show the theoretical estimates from eq.\ \ref{nusnut}.}
\label{Pr01}
\end{center}
\end{figure}

 Figs.\ \ref{Pr01} shows the dependence of the measured Nusselt numbers \nutt and \nust on the parameters \rast and R$_\rho$, for the case Le$=0.01$, Pr=0.1. The results show considerable scatter, as expected from the limited number of overturning times for which the simulations were run. Nevertheless, they appear consistent within about a factor of 2 with theoretical estimate (\ref{nusnut}).
 
 %numerically measured fluxes are shown and compared with the estimate (\ref{nusnut}), for Prandtl numbers 0.1 and 1 respectively, with Ra$_*$ and $R_{\rho}$ as parameters.  Viscosity (dominated by the lighter component of the plasma) is generally larger than, but of the same order of magnitude as, the diffusivity of the heavier component. The ratio, the Schmidt number $\mathrm{Sc}=\nu/ \ks=\mathrm{Pr/Le}$, is 10 in the results of Figs. \ref{Pr01},\ref{Pr1}. 

%In the low-Prandtl number results, the Nusselt numbers follow the expected relation within a factor 2. This is as close as could be expected given the scatter due to the stochastic nature of the flows and the limited number of overturning times (of the order 20) over which the simulations were run. The figure shows that the Nusselt numbers decrease with increasing density ratio, but this dependence disappears at the highest Rayleigh numbers. At Pr=1, the results still follow relation (\ref{nusnut}), but the effect of the density ratio on the Nusselt numbers is now much stronger, and remains present at the highest Rayleigh numbers covered by the simulations. Figs.\ \ref{rrhoPr1}, \ref{rrhoPr01} show this in a different representation of the results. 

%The results thus suggest that the dependence on the density ratio is restricted to Prandtl numbers of order unity and higher.

%\begin{figure}
%\begin{center}
%  \includegraphics[width=\hsize]{plots/NuS-NuT-1-01.pdf}
%  \caption{} 
%\label{Pr1}
%\end{center}
%\end{figure}

%\begin{figure}[h!]
%\begin{center}
%\includegraphics[width=\hsize]{plots/Rrho-NuT-1-01.pdf}
%\caption{Nu$_T$ as function of $R_{\rho}$ for $\rm Pr=1.0$ and $\rm Le=10^{-1}$. The thermal flux is %strongly affected by the stability parameter.}
%\label{rrhoPr1}
%\end{center}
%\end{figure}

\begin{figure}[h!]
\begin{center}
\includegraphics[width=\hsize]{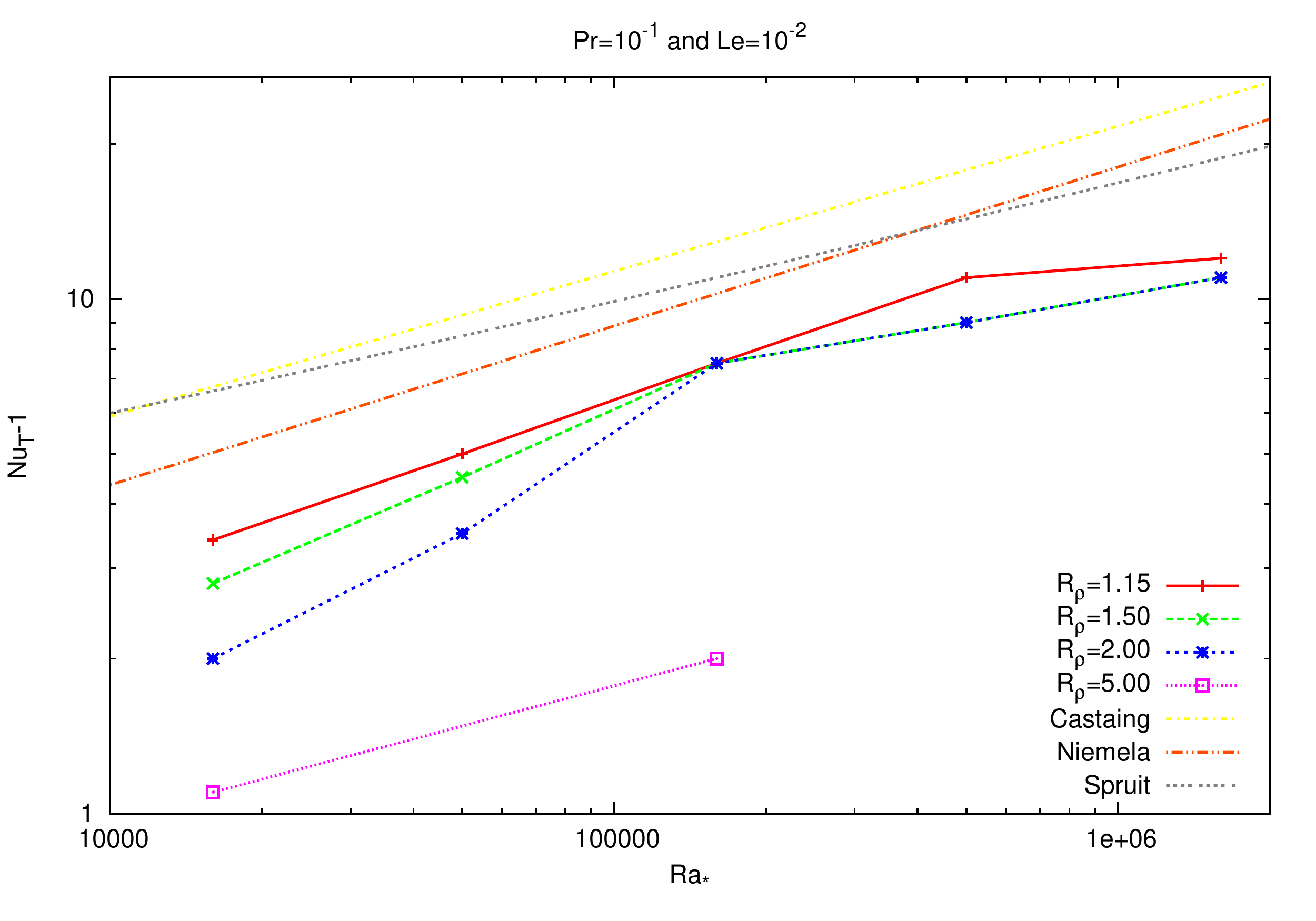}
\caption{Nu$_T$ as function of Ra$_*$ for $\rm Pr=10^{-1}$ and $\rm Le=10^{-2}$.}
\label{RaPr01}
\end{center}
\end{figure}

%
%\subsection{Dependence on {\rm Ra$_*$}}
%
The numerical results thus confirm that the {\em ratio} of thermal to solute transport does not depend much on parameters other than Lewis number and density ratio. The actual transport efficiency of both of course does depend on the Rayleigh number Ra$_*$. This is shown in Fig.\ \ref{RaPr01}, and compared with the thermal convection results discussed below in section \ref{mixing}.  In S92 an estimate of this relation was made for the asymptotic dependence at large Ra$_*$: $ \nut\approx 0.5\, \ras^{1/4}$ (for the essence of the derivation see section \ref{bound}). As above, the estimate can be made more accurate at lower Nusselt numbers by subtracting the diffusive flux:
\beq  \nut-1\approx 0.5\, \ras^{1/4}.\eeq
%Fig.\ \ref{RaPr01} shows that this simple estimate fits the results reasonably, for the range of Rayleigh numbers covered by the simulations. 

It is seen that this fits the numerical Nusselt numbers for density ratios near unity, but overestimates the heat flux at higher R$_\rho$. For the relation between \nust and \nutt this has little effect, since both are similarly affected by R$_\rho$. As discussed below, since the heat flux is fixed by the structure of the star, this uncertainty should not affect the expected mixing rate in a stellar interior much.

\begin{figure}[h!]
\begin{center}
\includegraphics[width=\hsize]{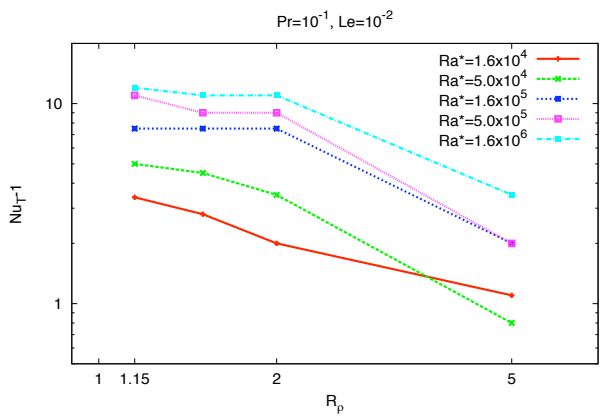}
\caption{Nu$_T$ as function of $R_{\rho}$ for $\rm Pr=10^{-1}$ and $\rm Le=10^{-2}$.}
\label{rrhoPr01}
\end{center}
\end{figure}

The same data is shown again in Fig.\ \ref{rrhoPr01}, with R$_\rho$ on the horizontal axis and \rast as parameter.

%\begin{figure}[h!]
%\begin{center}
%\includegraphics[width=\hsize]{plots/theor-Nu.pdf}
%\caption{Several power laws in the form Nu$_T \propto \alpha Ra^{\beta}_T$.}
%\label{fig:theor-Nu}
%\end{center}
%\end{figure}

%\begin{center}
%\includegraphics[width=\hsize]{plots/RaT-NuT-1-01.pdf}
%\caption{Nu$_T$ as function of Ra$_*$ for $\rm Pr=1.0$ and $\rm Le=10^{-1}$. Each
% vertical line represents a simulation set with (vertically decreasing) $R_{\rho}$. This means
% that dots at the top of a line have $R_{\rho}=1.15$ whereas dots at the bottom of a line have
% $R_{\rho}=2-5$. The theoretical diffusion corrected estimation by H. Spruit fits good for
% small values of Ra$_*$. }
%\label{fig:RaT-NuT2}
%\end{center}

\subsection{Dependence on aspect ratio}
\label{aspect}
The aspect ratio of $2:1$ used in the results reported above was tested against $5:1$ and $10:1$ at the same spatial resolution. For the reference simulation ($\rm Pr=0.1$, $\rm Le=0.01$, Ra$_*=10^5$, $R_{\rho}=1.15$) we find  Nu$_S=90$ and $\nut=8.5$. By comparison the simulation with $5:1$ results in $\nus=90$ and $\nut=9.0$. The most extended box with an aspect ratio of $10:1$ and a spatial resolution of $1500\times300$ has Nusselt numbers of  $\nus=80$ and $\nut=8.75$.  The aspect ratio thus has no significant influence on the dependences of the fluxes on input parameters in our simulations, within the fluctuations due to  the stochastic nature of the flow.

%\begin{figure}
%\begin{center}
%\includegraphics[width=9cm]{plots/NuT-Sc.pdf}
%\caption{Test of the impact of numerical diffusion, see text.}
%\label{fig:NuT-Sc}
%\end{center}
%\end{figure}

\subsection{Dependence on initial stratification}
\begin{figure}
\begin{center}
\includegraphics[width=0.45\hsize]{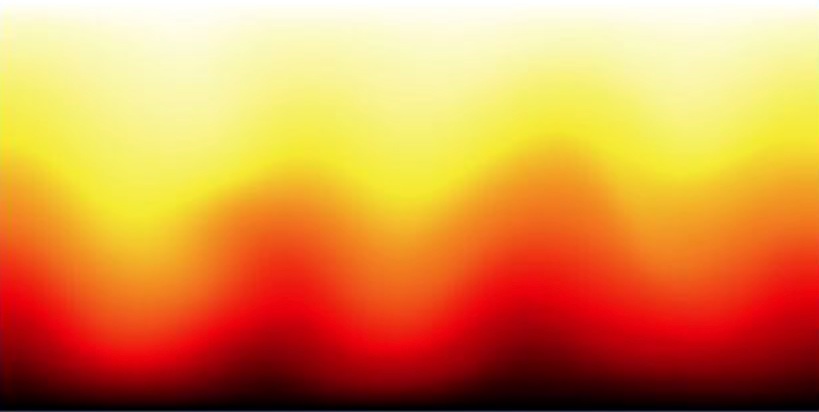}
\includegraphics[width=0.45\hsize]{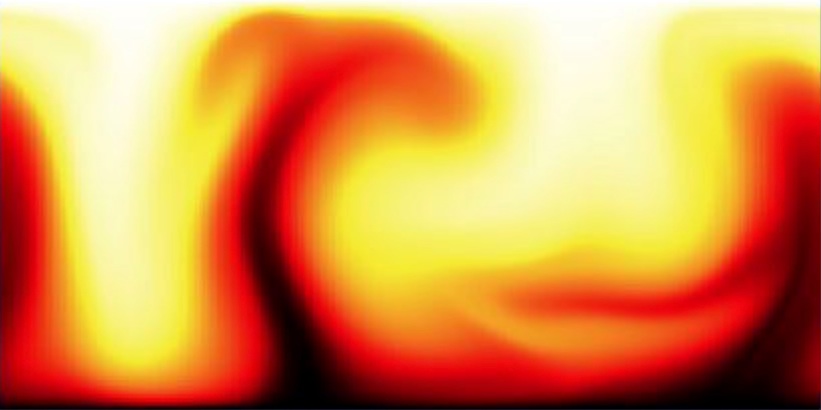}\\
\includegraphics[width=0.45\hsize]{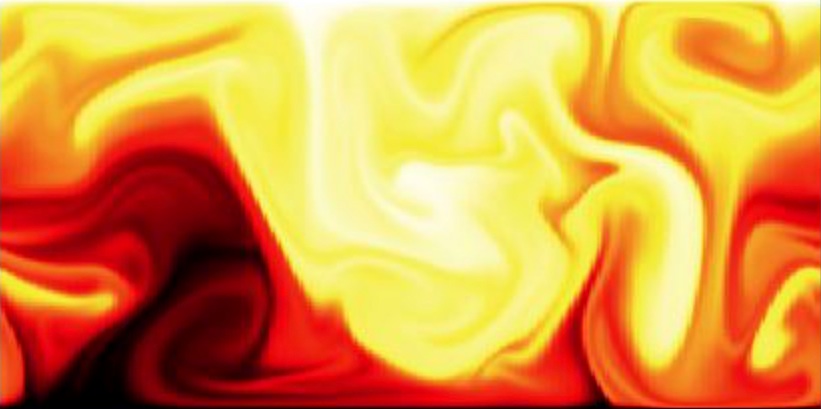}
\includegraphics[width=0.45\hsize]{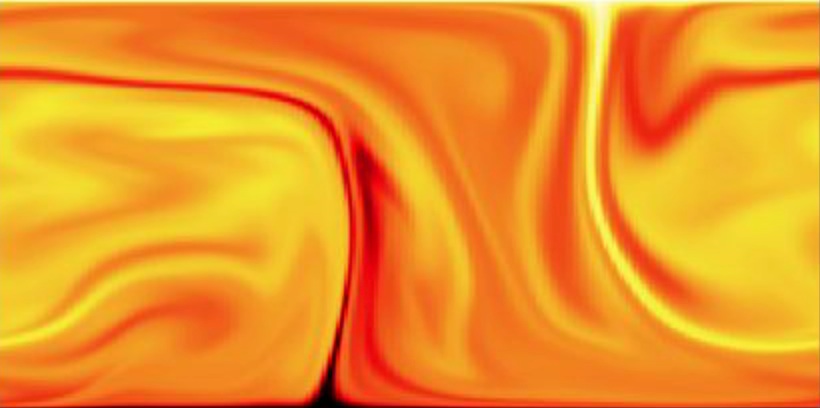}
\caption{Example of the development of an overturning flow from Kato oscillations. Time from left to right and top to bottom. See also movie at http://www.mpa-garching.mpg.de/$\sim$henk/movie.avi}
\label{kato}
\end{center}
\end{figure}

\begin{figure}
\begin{center}
\includegraphics[width=9cm]{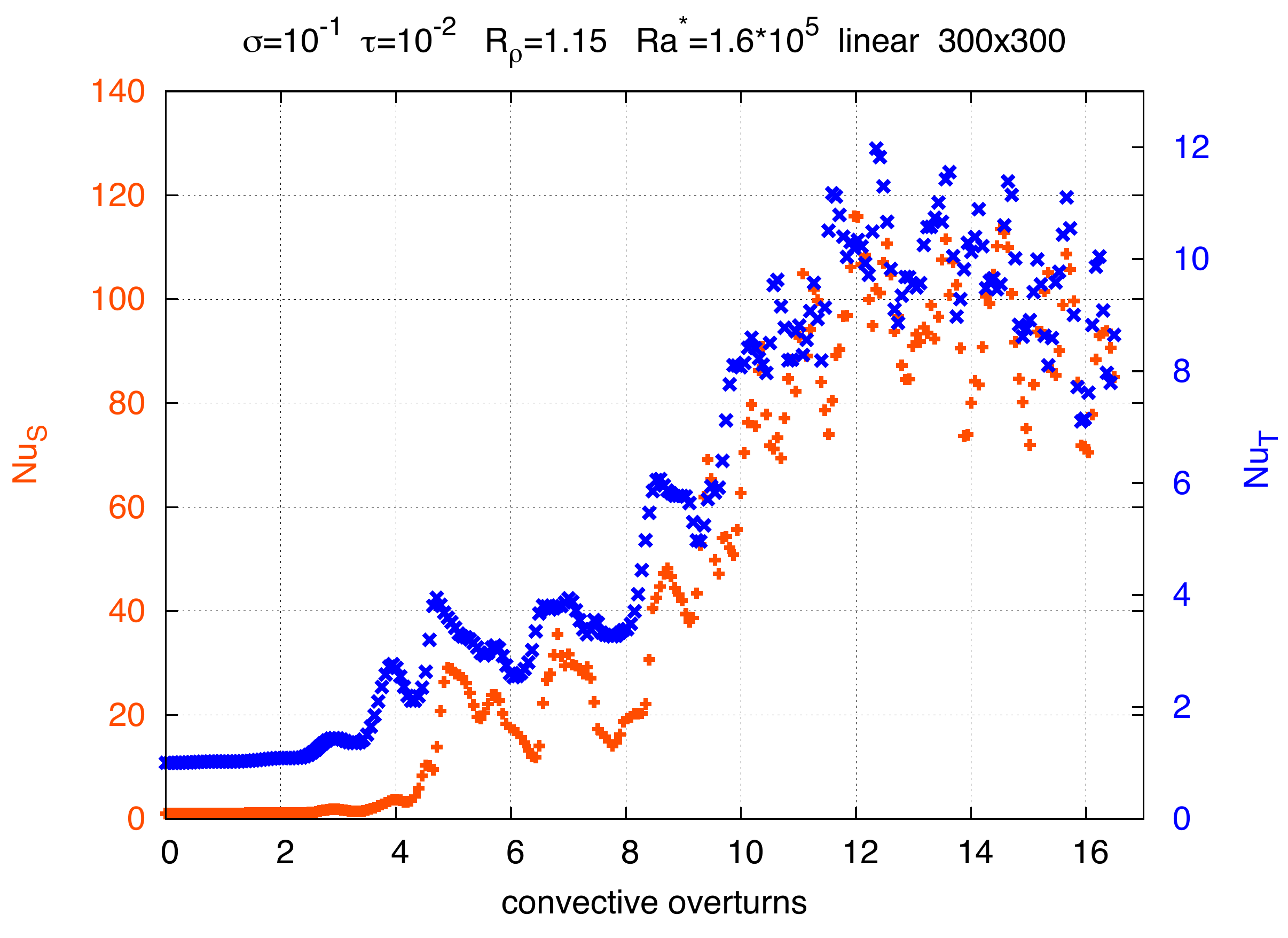}
\includegraphics[width=8.45cm]{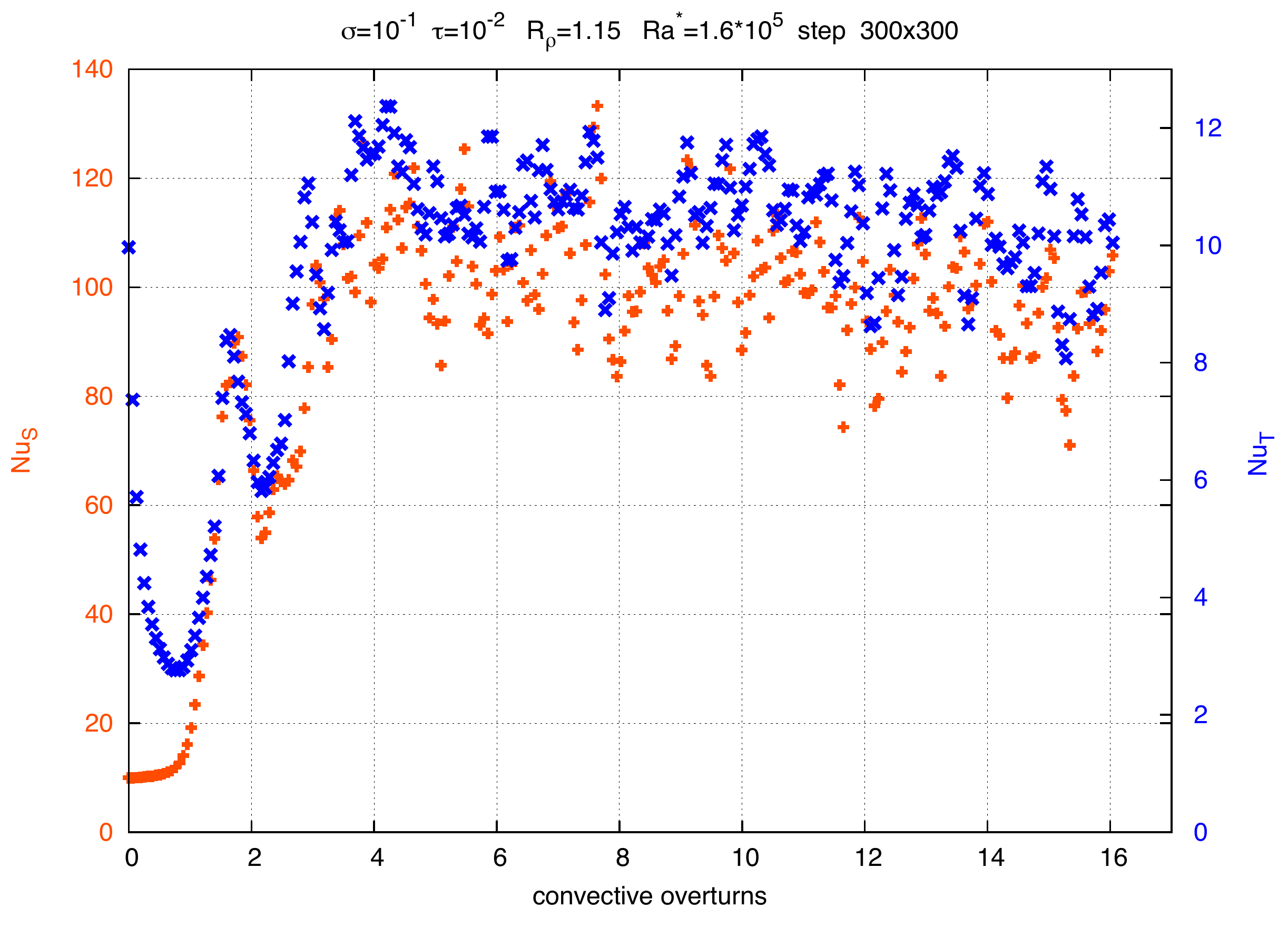}
\caption{Top panel: evolution of the Nusselt number for the linear initial stratification. Convective cells get established from Kato oscillations (cf. Fig.\ \ref{kato}) after about 10 turnover times. Starting the simulation from a step (bottom) saves computing time. At the end of the runs the Nusselt numbers of both simulations are the same within the statistical variations.}
\label{fig:evolution}
\end{center}
\end{figure}

As initial state we used either a constant linear gradient of temperature and solute between top and bottom values (`linear'), or a profile with boundary layers of width as estimated in \ref{bound} (`step'). The linear case shows how the oscillatory phase due to the Kato instability develops into overturning flow, see Figs.\ \ref{kato},\ref{fig:evolution}. The duration of the initial formation process is mainly determined by $R_{\rho}$ and Ra$_*$. The end state in both cases is statistically the same. The `step' as initial condition also covers cases that are stable in linear theory because of the subcriticality of the system, and can gain a large factor in computing time for small values in Ra$_*$ and high $R_{\rho}$. It was used in all the results reported above.

\subsection{Comparison with compressible results}
\label{compar}
The compressible simulations  are based on a  5th order weighted ENO scheme. 
Compressible fluids lead to restrictions in time stepping (due to the need to resolve sound waves). The compressible simulations take up to $100$ times longer for the same resolution compared to the incompressible solver. Therefore only a few tests have been done with the compressible code. The degree of compressibility is governed by the $\epsilon=d/H$ of layer thickness to pressure scale height; in the limit $\epsilon\rightarrow 0$ there is a direct translation between the compressible and the Boussinesq case (section \ref{Bouss}). The results of a numerical comparison with $\epsilon=0.1$ is shown in Table 1. The resulting Nusselt numbers do not differ significantly. 
%
%\begin{align}
%H_p & = p \Big ( \frac{\partial p}{\partial z} \Big ) ^{-1}  \approx \frac{p}{\rho g} \\
%\epsilon & = d/H_p
%\end{align}
%  
Simulations done with $\epsilon=1.0$ behave quite similar to these done with $\epsilon=0.1$. The mixing processes do not significantly differ as long as the Rayleigh number is high enough, $\mathrm{Ra} >5.0 \times 10^5$. 
At the present level of accuracy (a factor of 2, say), we conclude that the Boussinesq approximation gives the right results even for layer thicknesses approaching a scale height.
\begin{tiny}
\begin{table}[h]
\centering
\begin{tabular}{c | c | c | c | c | c | c | c}
Pr & Le & $R_{\rho}$ & $\ras$ & $\rm Nu_S^B$ & $\rm Nu_T^B$  & $\rm Nu_S^s$ & $\rm Nu_T^s$\\
\hline
\vspace{1\baselineskip}
$10^{-1}$& $10^{-2}$ & $2.0$ & $1.6 \times 10^5$ & $60$  & $8$ & $55$  & $7$  \\
$10^{-1}$& $10^{-2}$ & $1.2$ & $1.6 \times 10^5$ & $110$ & $12$ & $110$  & $11$  \\
$10^{-1}$& $10^{-2}$ & $2.0$ & $1.6 \times 10^6$ & $150$  & $12$ & $130$  & $10$  \\
$10^{-1}$& $10^{-2}$ & $1.2$ & $1.6 \times 10^6$ & $200$  & $16$ & $200$  & $14$ \\
$1.0$ & $10^{-1}$ & $2.0$ & $1.6 \times 10^5$ & $3.5$ & $2$ & $11$  & $1.5$  \\
$1.0$ & $10^{-1}$ & $1.2$ & $1.6 \times 10^5$ & $45$  & $15$ & $17$  & $5$  \\
$1.0$ & $10^{-1}$ & $2.0$ & $1.6 \times 10^6$ & $4$  & $3.5$ & $26$ & $10$  \\
$1.0$ & $10^{-1}$ & $1.2$ & $1.6 \times 10^6$ & $33$  & $11$ & $26$  & $10$  \\
\hline
\end{tabular}
\caption{Comparison of compressible and incompressible simulations. Thermal ($_{\rm T}$) and solute  ($_{\rm S}$) Nusselt numbers from the Boussinesq ($^{\rm B}$) and compressible ($^{\rm s}$) results. Layer thickness $d$ is 0.1 pressure scale height. }
\label{table-eps01-2}
\end{table}
\end{tiny}

\subsection{Multi-layer simulations}
\label{double}

\begin{figure}
\centering
\includegraphics[height=0.8\hsize]{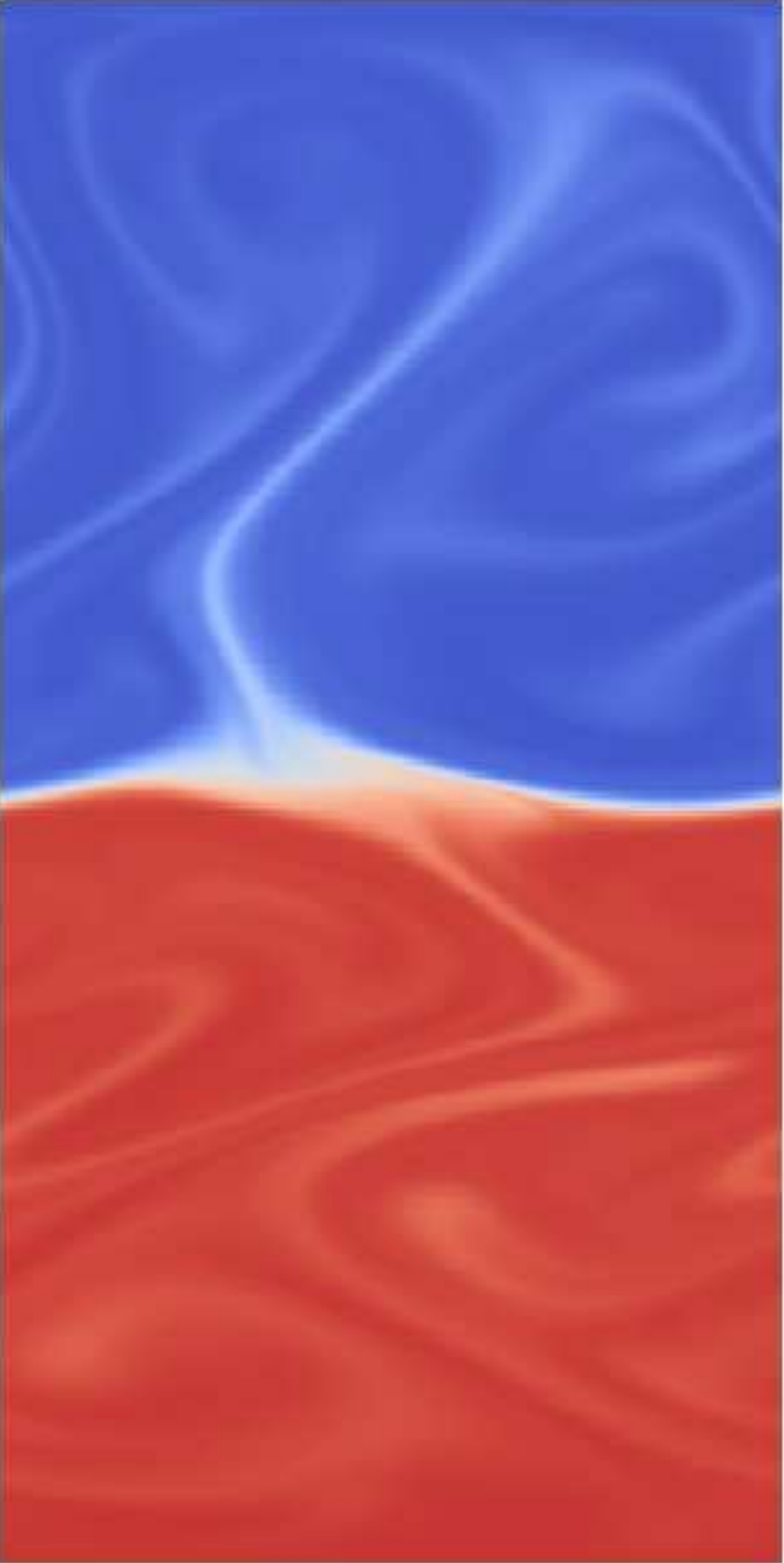}\hfil
\includegraphics[height=0.8\hsize]{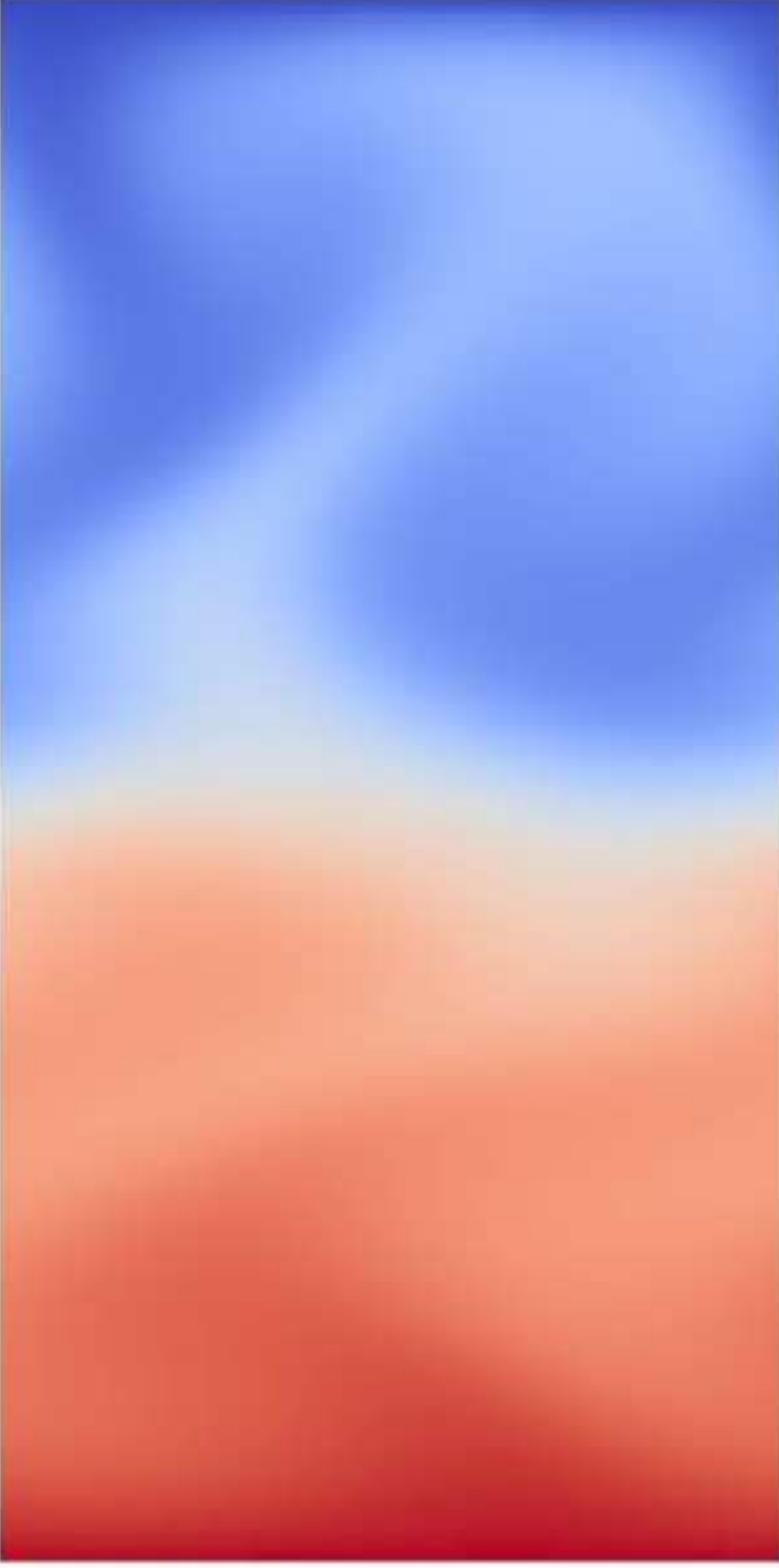}

\caption{Snapshot of a simulation including the free interface between two layers. left: solute, right: temperature. Pr=1.0, Le=0.01,$R_\rho$=1.15, Ra$_*=6\,10^5$. See movie at http://www.mpa-garching.mpg.de/$\sim$henk/double\_layer.avi } \label{fig:ds1}
\end{figure}

In all of the above we have assumed that the interfaces between the double diffusive layers can be approximated as solid boundaries. To test the reliability of this assumption, a few cases were run where the initial state consisted of two instead of a single step. In some, though not all of these runs, the division into two layers remained till the end. An example is shown 
in Fig.\ \ref{fig:ds1} for a case with $\rm Pr=1$, $\rm Le=10^{-2}$, $R_{\rho}=1.15$, Ra$_*=6\times 10^5$ and a resolution of $500\times 500$. Note the approximate (anti-)symmetry of the plumes near the interface in the middle, a phenomenon known from laboratory experiments. It is caused by the continuity of the horizontal velocity across the interface enforced by viscosity. This symmetry becomes less marked in simulations at lower Prandtl numbers.

\begin{figure}
\centering
\includegraphics[width=\hsize]{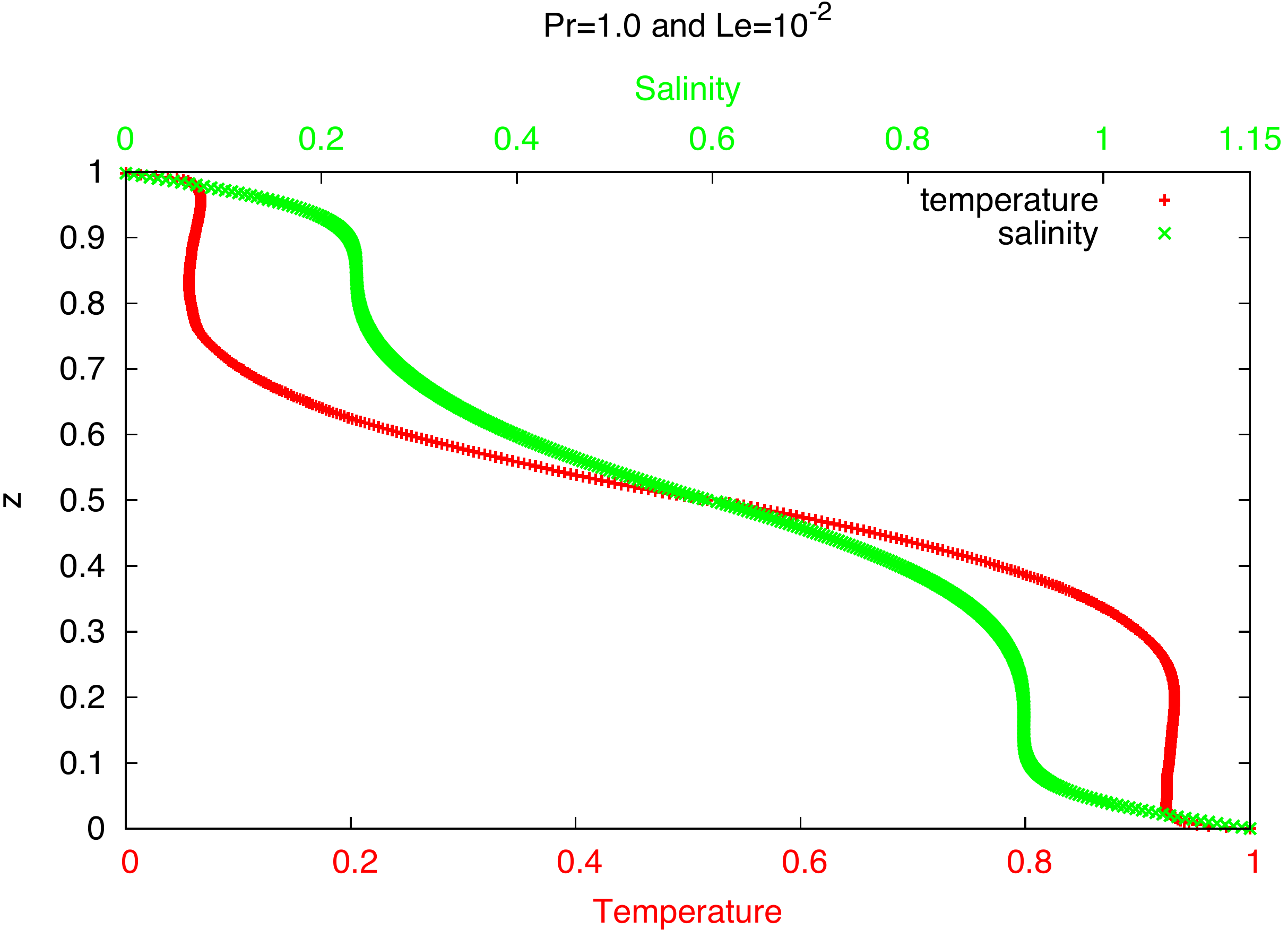}\hfil
\caption{Average temperature and solute profiles with height $z$ in the double-layer simulation of Fig.\ \ref{fig:ds1}.} \label{dsmean}
\end{figure}

Fig.\ \ref{dsmean} shows horizontal and temporal average profile of temperature and solute with height. The transition in the middle is  broad compared with the boundary layers at top and bottom. Inspection of the time dependent flow shows that this is due to two separate effects. One is the displacements of the interface by surface waves, which smoothen the average gradient without changing the actual fluxes across the interface. In addition there is a real mixing effect associated with breaking of the surface waves, but it remains localized around the interface. So as to maintain the same heat flux, broadening by this mixing has to be compensated by a larger amplitude of the transition. The effect is equivalent to a decrease of the coefficient $a$ in (\ref{pow}), and a decrease of the effective diffusion rate.

%\newpage

%\newpage

\section{Application to stars}
\label{appli}
\subsection{The low-{\rm Pr}, low-{\rm Le} limit}
The astrophysical limit, where the thermal diffusivity is orders of magnitude larger than the other diffusivities, actually simplifies the physics of the double diffusive problem.
As predicted by relation (\ref{nusnut}) and in agreement with our numerical results, the solute flux is low compared with the thermal flux carried by the flow, in this limit. The amount of solute which is in the process of being transported from the lower to the upper boundary is therefore small at any point in time. The effect of its  buoyancy on the dynamics of the flow is then only a perturbation. 

To the extent that the solute and viscous boundary layers are thin compared with the thermal boundary layer, the flow the interface between layers can be approximated as a sharp jump. The position of the interface fluctuates due to the presence of surface waves, but these waves do not transport a net flux (averaged over a wave period, say). 

The effect of the interfaces is thus equivalent to those of solid boundaries, and the flow inside the layer is almost the same as thermal convection between plates at  the same temperature difference.

Because of this useful circumstance, the heat flux under the low-Pr, low-Le astrophysical conditions can be found from the equivalent thermal convection problem, and the solute flux then follows simply from relation (\ref{nusnut}). 
For this equivalent convection problem, laboratory results and analytical estimates are available, and these can then be used to extrapolate the numerical results to higher Rayleigh numbers.

\subsection{Extrapolation}
\label{mixing}
Convection experiments in gaseous Helium at temperatures just above the critical point by Castaing et al (1989) yielded the fit
\begin{equation}
{\rm Nu_T}=0.23\ {\rm Ra}^{0.282}. 
\label{NuTCastaing}
\end{equation}
over the range  $10^7<\mathrm{Ra}<10^{14}$.
More recent measurements by Niemela et al. (2000) using superfluid $^3$He, over the remarkable range $10^6 \leq {\rm Ra} \leq 10^{17}$ gave a marginally different result:
\begin{equation}
{\rm Nu_T}=0.124\, {\rm Ra}^{0.309}.
\label{NuTNiemela}
\end{equation}
This can be compared with the estimate based on a 2-dimensional argument in S92:
\beq \nut \approx 0.5\, \mathrm{Ra}_*^{0.25}.\label{s92} \eeq
Since the Prandtl number in the laboratory experiments (of order 0.7) is not far from unity, Ra in (\ref{NuTCastaing}, \ref{NuTNiemela}) can be identified with \rast in (\ref{s92}).  The expressions above are of the form ${\rm Nu_T}=a\, {\rm Ra}^\beta$. As in the above, we subtract 1 from the Nusselt number, for a marginally better fit at low \nutt:

\beq {\rm Nu_T}-1=a\, {\rm Ra_*^\beta}.\label{pow} \eeq
To make these things astrophysically useful, we need to express \rast in terms of the superadiabatic gradient (eq.\ \ref{ras}):
\beq {\rm Ra}_*=(\nabla-\nabla_{\rm a})R\,\epsilon^4,\label{eq0}\eeq
where
\beq R={g H^3/\kappa_{\rm T}^2}\label{RR}\eeq
is a function the local thermodynamic state, but not of the (still unknown) temperature gradient, and $\epsilon=d/H$ is the double diffusive layer thickness $d$ in units of the pressure scale height $H$. 

In the limit $\epsilon\rightarrow 0$, this Rayleigh number is equivalent to that in a Boussinesq  calculation. To translate the corresponding Nusselt number into an astrophysical flux, however, the complication discussed in section \ref{Bouss} has to be taken into account. As derived there, the identification (eq. \ref{ident}) has to be made in terms of the ratio $f$ of convective flux $F_{\rm c}$ to the part $F_{\rm rs}$ of the radiative flux that is carried by (only) the superadiabatic part of the temperature gradient. In terms of the logarithmic gradients, this ratio is:
\beq 
f^{\rm s}=F_{\rm c}/F_{\rm rs}= {F-F_{\rm r}\over F_{\rm rs}}={\nabla_{\rm r}-\nabla\over\nabla-\nabla_{\rm a}} \label{fs}.
\eeq
With (\ref{pow},\ref{eq0},\ref{RR}) this yields:
\beq\nut -1= f^{\rm s} = (\nabla_{\rm r}-\nabla_{\rm a}){R\over {\rm Ra}_*}\epsilon^4-1= a\, {\mathrm{Ra}}_*^\beta.\label{eq2}\eeq
The last equality defines a relation between the layer thickness $d$ and Ra$_*$, under the assumption of a Nusselt number of the form (\ref{pow}). With (\ref{ras}), this then yields the superadiabaticity as a function of $d$. 

\begin{figure}
\begin{center}
\includegraphics[width=\hsize]{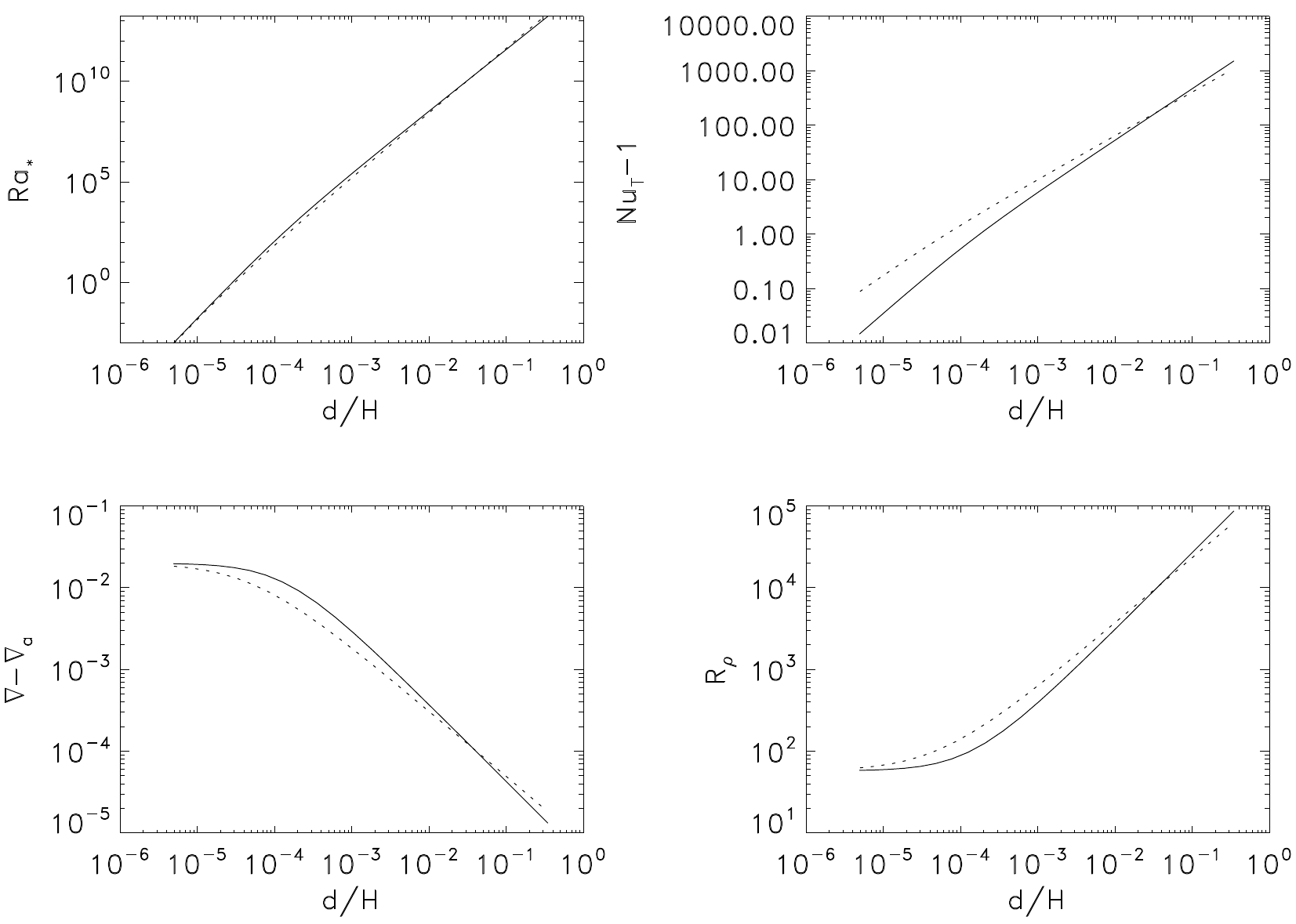}
\caption{Semiconvection for the parameters of a 15 $M_\odot$ MS star. Top left: modified Rayleigh number as a function of the double-diffusive layer thickness $d$ (eq. \ref{eq2}), right: Nusselt number, bottom left: the corresponding superadiabaticity, right:  density ratio. Solid line: using the fit (\ref{NuTNiemela}), dotted:  fit (\ref{s92}). The numerical results shown in Fig.\ \ref{RaPr01} lie just below the solid line in the top right panel, between $d/H=10^{-4}$ and $d/H=10^{-3}$.}
\label{compa}
\end{center}
\end{figure}

The mixing rate, as function of the layering thickness $d$, follows from eqs. (\ref{nusnut}, \ref{pow}). Introducing the effective solute diffusivity $\kappa_{\rm Se}$:
\beq \kappa_{\rm Se}/\ks=\nus  = [1+(\kt/\ks)^{1/2}/R_\rho]\,(\nut -1)+1,\label{mix}\eeq
where (eq. \ref{rrho}):
\beq R_\rho=-N_\mu^2/N_{\rm T}^2={g\,{\rm d}\ln\mu/{\rm d}z\over g(\nabla-\nabla_{\rm a})/H}= {\nabla_\mu\over\nabla-\nabla_{\rm a}}.\eeq

In the limit $\nus\gg 1$, $\nut \gg 1$, and assuming the second term in the first bracket to dominate, this reduces to:
\beq 
\kappa_{\rm Se}/\ks=\nus  = ({\kt\over \ks})^{1/2}{\nabla_{\rm r}-\nabla_{\rm a}\over\nabla_\mu}.\label{mix}\eeq
This is the same as derived in S92, except that the contribution of radiation pressure in the equation of state was also taken into account there, which yielded:
\beq 
\kappa_{\rm Se} = ({\kt \ks})^{1/2}\left({4\over\beta}-3\right){\nabla_{\rm r}-\nabla_{\rm a}\over\nabla_\mu},\label{mixb}\eeq
where $\beta=P_{\rm g}/(P_{\rm g}+P_{\rm r})$ is the ratio of gas pressure to total pressure.

\subsection{A 15 $M_\odot$ star}
\label{stars}
We are now in the position to estimate  the range of parameter values for semiconvection in a star, and which part of this range is covered directly by numerical simulations. Consider  the important case of massive stars around main sequence turnoff. We use a model  kindly provided by A.~Weiss (model `fzm15\_151'). Characteristic values for the physical quantities in the semiconvective zone this model 
are $g\approx 10^6$ cm/s$^2$, $\ks \approx 1$, $\kt \approx 3\,10^8$ cm$^2$/s, $H\approx 2\,10^{10}$ cm, $\nabla_{\rm a}=0.4$, $\nabla_{\rm r}-\nabla_{\rm a}\approx 0.02$, $\nabla_\mu\approx 1$. Fig \ref{compa} shows the dependence of Ra$_*$, Nu$_{\rm T}$,  $\nabla-\nabla_{\rm a}$ and $R_\rho$ on $\epsilon=d/H$ as determined from (\ref{eq2}), for two of the power laws above. The lower left panel shows how the superadiabaticity increases with the decrease of convective efficiency at small layer thickness. At thicknesses less than $10^{-4}$ convection becomes inefficient at transporting heat, and $\nabla$ saturates to its radiative value $\nabla_{\rm r}$. The increase of the density ratio mirrors this: because of the efficiency of convection  at high layer thickness and the resulting weak superadiabaticity it reaches very high values.

The effective He-diffusivity from (\ref{mixb}) is $ \kappa_{\rm Se}\approx 10^3$ cm$^2$/s. The mixing time scale over a pressure scale height is thus about $10^{10}$ yr.

\subsection{Discussion}
\label{discussion}

As in the case of ordinary convection, there are far fewer intrinsic parameters in semiconvection than physical quantities defining the physical state in a stellar interior. This allows a significant volume of astrophysically relevant parameter space to be covered by a grid of numerical simulations. The low value of the viscosity and constituent diffusivity compared with the thermal diffusivity present a limiting case that actually simplifies the double diffusive problem greatly. In this limit the results become nearly independent of the Prandtl number. The remaining parameters can be represented by the dimensionless quantities $R$ (eq. \ref{RR}) for the thermodynamic state of the plasma, the Lewis number $\ks/\kt$, the density ratio $R_\rho$ and the layer thickness $d/H$ (eqs.\ \ref{eq2},\ref{mix}).  

The simulations were all done in 2-D, so tests in 3-D will be needed for verification. It is  unlikely, however, that the results will turn out very different,  at least within an astrophysical factor of 2. The reason is that in the low-Pr, low-Le limit the flow in the layers is almost equivalent to ordinary convection between plates. Due to the low solute diffusivity, the amount of solute in the bulk of the layer is small. It can then be treated as a perturbation, as assumed  in S92. Known laboratory results for convection at very high Rayleigh numbers can then be used to extrapolate the numerical results. As shown in section \ref{appli} (Fig.\ \ref{compa}), this makes predictions similar to the simple model used in S92. In particular the effective mixing rate is very low.

Stochastic fluctuations in the flow produce scatter in the fluxes of heat and solute, which affect the measured averages (cf.\ Table 1). The accuracy of these averages could be improved with longer runs.

Most of the results presented are based on simulation of a single double diffusive layer. As we found in \ref{double}, this does not capture broadening of the interfaces between layers by breaking surface waves (see movie at Fig.\ \ref{fig:ds1}). Since shallower gradients imply lower fluxes, such a broadening process reduces the effective mixing rate.

The physics determining the thickness of the individual double diffusive layers remains uncertain. In the equivalent geophysical examples semiconvective zones always consist of many very long-lived layers.  In the absence of a good theory for the layer thickness and its evolution it is not possible to translate this finding to an astrophysical setting, however. 

Observations in the east-african rift lakes (e.g. Schmid et al. 2010) show that layers first forming at the boundary of an expanding double-diffusive zone are always thin, subsequently growing slowly by a process of merging. This is likely to happen in a growing stellar semiconvective zone as well. Unclear, however, is how many layers will be left at the end of the semiconvective phase. If only a few are left, the location of the boundaries between these layers may well vary somewhat randomly between stars. This would introduce a stochastic element in the internal structure that might be related to observed variations in evolution tracks.

\subsection{Conclusions}

The results show that the behavior of double diffusive convection as observed in geophysics and the laboratory applies also to the astrophysical case of low Prandtl and Lewis numbers. In particular, the process takes place in the form of the  characteristic double diffusive layering known and theoretically understood since the 1980's. This is in contrast with the models of linear or nonlinear oscillations on which most recipes used in stellar evolution are based. 
The results show that the well known `square root' relation between the solute flux and the thermal flux (\ref{nusnut}) observed in laboratory experiments, holds also in the astrophysical case of high thermal diffusivity. This relation is the most important factor determining the effective mixing rate.
  
By a direct comparison of Boussinesq and fully compressible simulations we find that the two give equivalent results in the limit of small layer thickness. The Boussinesq calculations even capture the essence for layers as thick as a scale height. 

We find that in the limit of low Prandtl number the results become nearly independent of Pr. This greatly reduces the parameter space to be covered by the grid of calculations. 

On the basis of these results we give fitting formulas from which  superadiabaticity and mixing rate can be calculated as functions of the physical parameters $\nabla_r$, $\nabla_a$, $R=gH^3/\kt^2$, Le$=\ks/\kt$ and layer thickness $d$. For the stellar model used for illustration, however, the effective mixing rate is essentially negligible.

\subsection{Acknowledgments}
F.\ Zaussinger is grateful to F.\ Kupka and H.J.\ Muthsam for discussions 
on the numerical solution of binary mixture equations with high resolution methods. He was 
supported by the DFG within the project `Modelling of
diffusive and double-diffusive convection' (projects KU 1954-3/1 and KU 1954-3/2 in SPP 1276/1 and SPP 1276/2, project leader F.\ Kupka) within the interdisciplinary Metstroem project. He was also supported by the Austrian Science 
Foundation (project P20973, Numerical Modelling of Semiconvection, project leader 
H.J.\ Muthsam). We also thank H.\ Grimm-Strele for the implementation of a parallel Poisson solver in the ANTARES code suite.

%\bibliographystyle{plain}
%\bibliography{fzbib}

\end{document}